\documentclass[12pt,a4paper,final]{iopart}

\usepackage{iopams}  
\usepackage{graphicx}
\usepackage[breaklinks=true,colorlinks=true,linkcolor=blue,urlcolor=blue,citecolor=blue]{hyperref}

\begin{document}

\title[Surface-electrode trap with an integrated permanent magnet]{Surface-electrode trap with an integrated permanent magnet for generating a magnetic-field gradient at trapped ions}

\author{Yuji Kawai$^{1}$, Kenji Shimizu$^{1}$, Atsushi Noguchi$^{2}$, Shinji Urabe$^{1}$, and Utako Tanaka$^{1}$}
\address{$^1$Graduate School of Engineering Science, Osaka University,1-3 Machikaneyama-cho, Toyonaka, Osaka 560-8531, Japan}
\address{$^2$Research Center for Advanced Science and Technology (RCAST),
The University of Tokyo, Meguro-ku, Tokyo, 153-8904, Japan}
\ead{k.shimizu@griffith.edu.au}

\begin{abstract}
We report on a surface-electrode trap with SmCo magnets arranged in a quadrupole configuration underneath the trap electrode. Because the distance between the magnets and the trapped ions can be as little as several hundred micrometers, a large magnetic field is produced without any heat management. The magnetic-field gradient was measured using the Zeeman splitting of a single trapped $^{40}$Ca$^+$ ion at several positions, and a field gradient of 36 T/m was obtained. Such a field gradient is useful for the generation of a state-dependent force, which is important for quantum simulation and/or quantum gate operation using radio-frequency or microwave radiation.

\end{abstract}

\vspace{2pc}
\noindent{\it Keywords}: ion trap, magnetic-field gradient, quantum simulation
\vspace{17pc}

\section{Introduction}

Trapped atomic ions have been regarded as a very promising physical system in quantum information processing. They have been used for many proof-of-principle experiments such as fundamental quantum gate operations\cite{NIST1995, Blatt2003cnot, Blatt2006cnot}, generation of entangled states\cite{NIST1998twoent, NIST2000fourent, Blatt2004}, and quantum simulation of coupled spins\cite{Schaetz2008}. In these pioneering studies, the quantized motional and internal states of trapped ions were controlled using laser radiation. In spite of such successful works, the use of laser radiation for coherent manipulation has a few issues. If two levels are coupled by a Raman transition, spontaneous emission causes the destruction of quantum coherence\cite{NISTspem2007}. On the other hand, if a ground state and a metastable state are chosen as a two-level system and driven by a quadrupole transition, hertz-level laser stability is typically demanded. In order to achieve further scalability, it is desirable to develop a system that is not affected by the decoherence due to spontaneous emission and is less demanding of the laser system.

	The use of radio frequency (rf) or microwave for the manipulation of quantum states has several advantages from these viewpoints\cite{Wunde2001, Chia2008}. Spontaneous emission does not affect the gate operation, which would lead to better fidelity in quantum-state manipulation. In addition, the generation and control of rf or microwave radiation requires a simpler system compared to that of laser radiation. Thus, a laser system is used only for laser cooling and state detection, which reduces the demand on their stability. To perform the quantum simulation of interacting spins or two-ion entangling gates, the state coupling of motional and internal quantum states is required. As proposed in \cite{Wunde2001}, exposing an ion string to a magnetic field with spatially varying magnitudes enables such coupling. The coupling strength is proportional to the square of the field gradient.

	Several studies aimed at generating a magnetic-field gradient at trapped ions have been reported. Individual addressing of trapped ions and coupling of motional and internal states have been performed using permanent magnets, which were set at the outside of end electrodes of a linear Paul trap \cite{Wunde2009, Wunde2012}. However, the magnitude of the field gradient is limited by the distance between the ion and magnets because it is generally difficult to integrate magnets inside a conventional linear Paul trap. One approach involved a current introduced on wires fabricated in a surface-electrode trap for the generation of magnetic-field gradients\cite{Wang2009, Wunde2013}, which can reduce the distance between the ions. This design is suitable for improving scalability; however, it requires heat management. Another approach involved the use of oscillating magnetic fields in a surface-electrode trap\cite{Osp2008, Osp2011, NISTtech2013, Osp2014,  Oxfordmw2014}. The use of the near-field amplitude gradient produced in a surface-electrode trap was proposed \cite{Osp2008}, and gate operation was demonstrated  \cite{Osp2011}. 

	Here, we present a different approach to generate a large field gradient at trapped ions. We integrate permanent magnets underneath a surface-electrode trap. This structure remarkably reduces the distance between permanent magnets and trapped ions. A large field gradient can be expected without any heat management. We discuss design considerations and present the numerical results of the magnetic field at the trapping region. We show that the trap is implemented by assembling a multi-segmented surface-electrode trap and an alumina plate in which permanent magnets are buried. To evaluate the field gradient, we observe the Zeeman splitting of the $^2S_{1/2}$ - $^2D_{5/2}$ transition in a $^{40}$Ca$^+$ ion at several positions. Finally, we discuss the generation of an even larger field gradient.

\section{Trap design}

The trap is composed of two parts; a surface-electrode trap and a magnet layer. The surface-electrode trap is made of a square alumina substrate, the surface of which is gold-plated to form electrodes. The magnet layer is also made of a square alumina plate, which has holes for placing permanent magnets. Both the trap electrode and the magnet layer are squares of a side 11.5 mm. The magnet layer is glued underneath the trap with a conductive epoxy to make these two squares overlap. Then, the trap is glued on a ceramic pin grid array (CPGA) mount. 

\vspace{3pc}

\begin{figure}[htb!]
 \centering
 \includegraphics[width=70mm, trim=0mm 0mm 0mm 0mm, clip]{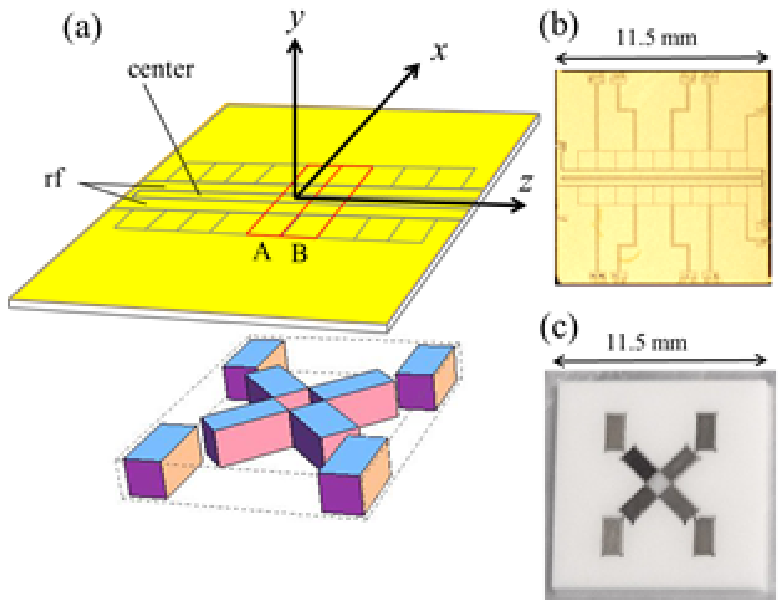}
 \caption{\label{figureone}Structure of trap electrode with permanent magnets. (a) Surface-electrode trap and magnet layer. Ions are trapped above the surface along the $z$ direction. Though the trap and the magnets are shown separately in this drawing for clarity, the eight permanent magnets are attached underneath the trap. The labels A and B represent the trapping regions discussed in this work (See text). (b) Photograph of the trap chip. (c) Photograph of the magnet layer placed underneath the trap chip (b).}
\end{figure}

\subsection{Segmented surface-electrode trap}

The trap design is based on a segmented linear surface-electrode trap (Fig. \ref{figureone}). An rf voltage is applied to the two rf electrodes for radial confinement. A direct current (dc) electrode called center electrode is placed between the rf electrodes, and it is used for adjusting the height of the static potential minimum. There are nine dc electrodes on either side of the rf electrodes for axial confinement. Ions can be trapped in any region except for the two outer regions, and they can also be shuttled between these regions by controlling the applied dc voltages. In the present work, we focus on two regions labeled A and B. These regions have different magnetic-field gradients when the magnet layer is attached. 

    The trap substrate is an alumina plate of area 11.5 mm$^2$ square and thickness 0.2 mm. Electrodes are formed through gold plating using Ti and Pd as adhesive layers. The thickness of the gold electrode is approximately 5 $\mu$m. The widths of the rf electrodes and center electrode are 300 $\mu$m and 100 $\mu$m, respectively. All the dc electrodes for axial confinement are squares of dimensions 1 mm by 1 mm. The spacing around the rf electrodes is 50 $\mu$m, while the other spacing is 25$\mu$m.

\subsection{Magnet layer}

Eight rectangular parallelepiped magnets are located underneath the trap electrode. We choose the quantized axis in the $x$ direction and the trap axis in the $z$ direction. Our requirement is a large gradient of the $x$ component of magnetic field along the $z$ direction. To realize this, four of the magnets are aligned in the quadrupole configuration at the center of the layer. Ideally, at $x=0$, no components exist along the $y$ and $z$ directions. Further, only the magnitudes of $x$ components exist and cross zero at $z=0$. The large magnetic field causes a large detuning in cooling laser frequency; therefore, we place four other magnets in order to reduce the magnitude of the magnetic field outside the quadrupole configuration. In addition, these four magnets enable experiments under different field-gradient conditions. SmCo magnets with dimensions of 1 mm $\times$ 1 mm $\times$ 2 mm are fit in a pattern made of alumina with a thickness of 1 mm, as shown in Fig. \ref{figureone} (c). The magnet layer is glued underneath the trap chip. 

\begin{figure}[htb!]
 \centering
 \includegraphics[width=70mm, trim=0mm 0mm 0mm 0mm, clip]{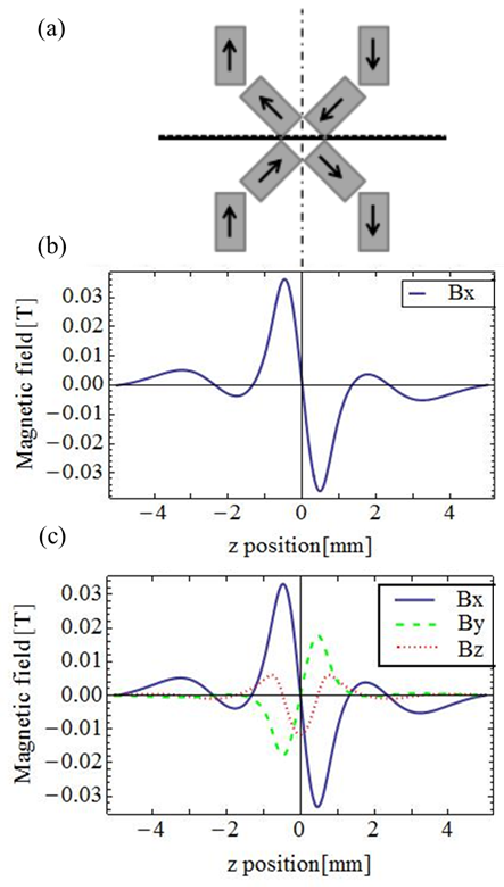}
 \caption{\label{figuretwo}Results of calculation of magnetic field generated by the magnet layer. (a) Configuration of eight magnets. The arrows indicate magnetization vectors inside the magnets. (b) Magnetic field along the $z$ direction (horizontal line) at $y$=150 $\mu$m, which is 350 $\mu$m from the top surface of the magnetic layer. In this ideal case, i.e., when the trap and the magnet layer completely overlap, both $y$ and $z$ components are zero. (c) Same calculation as (b), with a shift between the trap and the magnet layer of 100 $\mu$m assumed in the $x$ direction; that is, ions are located along the dotted horizontal line in (a) and 350 $\mu$m above the magnet-layer surface. In this case, ${B_y}$ and ${B_z}$ are not zero.}
\end{figure}

    For a quantitative estimation, we numerically calculated the magnetic field using Radia software package \cite{Radia} for the configuration of magnets shown in Fig. \ref{figuretwo}(a). Figure \ref{figuretwo}(b) shows the $x$ component of the magnetic field along the $z$ direction (horizontal line in Fig.\ref{figuretwo}(a)) 350 $\mu$m away from the top surface of the magnet layer. Owing to the symmetry of quadrupole configuration, both ${B_y}$ and ${B_z}$ are equal to zero in the ideal case. Because of the existence of the outer four magnets, the magnetic field is reduced in the outer region. The maximum field gradient is obtained at $z=0$, and a smaller gradient is obtained around the other zero-crossing points.  

	Because the trap chip and the magnet layer are separated parts originally, a shift between two squares may exist when they are glued, which results in the generation of an unwanted magnetic field at the trapped ions. Figure \ref{figuretwo} (c) shows a calculation similar to that in (b) but with a shift between the trap chip and the magnetic layer of 100 $\mu$m, as shown by the dotted horizontal line in Fig. \ref{figuretwo}(a). At $z=0$, ${B_z}$ becomes large, which changes the direction of the quantized axis. ${B_y}$ also arises around $z=\pm0.5 $mm. In the region $\mid z \mid> $1 mm, the unwanted magnetic-field components are close to zero.

　\smallskip 

\vspace{1pc}

\begin{figure}[htb!]
 \centering
 \includegraphics[width=70mm, trim=0mm 0mm 0mm 0mm, clip]{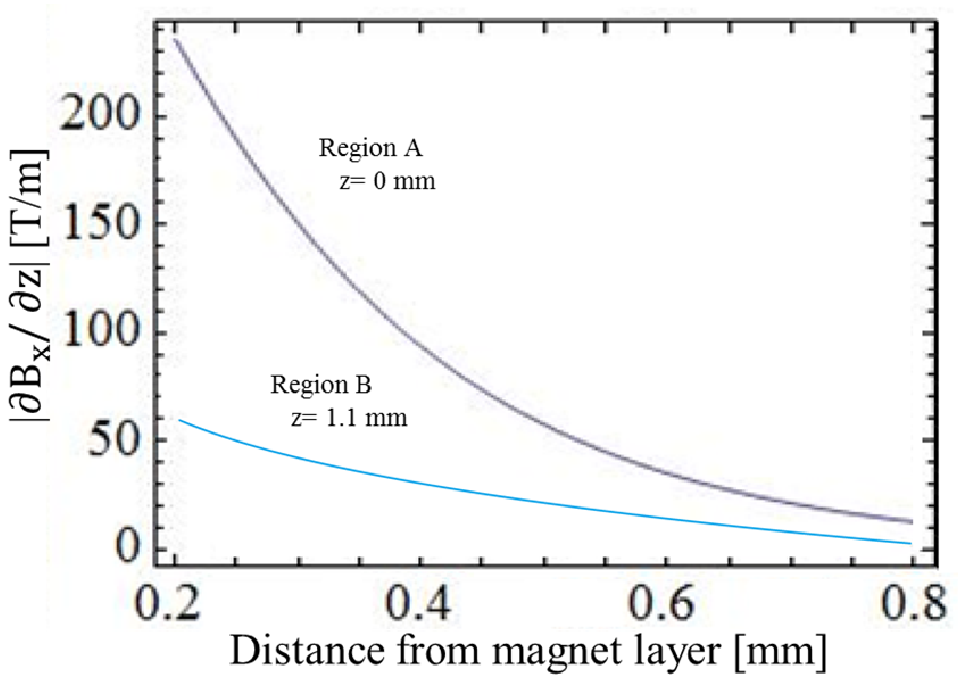}
 \caption{\label{figurethree}Magnitude of the magnetic-field gradient versus the distance between the ion and the top surface of the magnet layer for trapping regions A and B.}
\end{figure}

The dependence of the calculated magnetic-field gradient on the distance between the trapped ion and the top surface of the magnet layer is shown in Fig. \ref{figurethree} for the different trapping regions. When an ion is trapped in region B at a height of 350 $\mu$m from the magnet layer, a magnetic-field gradient greater than 100 T/m is estimated. Likewise, according to Fig. \ref{figurethree}, when an ion is confined in region B at a height of 350 $\mu$m, a magnetic-field gradient of approximately 38 T/m is estimated.

\section{Experimental Setup}
The trap on the CPGA mount is placed in a vacuum chamber with a vacuum level of 10$^{-8}$ Pa. Calcium ions are loaded by photo-ionization with a 423-nm laser for the $^1S_0$-$^1P_1$ transition and a 375-nm laser for ionization. Doppler cooling is conducted by the $^2S_{1/2}$-$^2P_{1/2}$ transition at 397 nm with an 866-nm laser for pumping back the ion from the $^2D_{3/2}$ state to the $^2P_{1/2}$ state. The Zeeman splitting is measured by using the quadrupole transition between the $^2S_{1/2}$ and $^2D_{5/2}$ states at 729 nm with a quenching laser connecting the $^2D_{5/2}$ and $^2P_{3/2}$ states at 854 nm.  All of the lasers for photo-ionization, Doppler cooling, and quenching are introduced in the $z$ direction. The 729-nm laser is introduced from the opposite side in the $z$ direction. The polarization of the 729-nm laser is set to be in the $x$ direction so that only the $\Delta m_J=\pm 1$ transitions occur according to the selection rule \cite{James1998, Roos2000}. Coils exist for the compensation of unwanted magnetic field outside the chamber. The fluorescence of ions is collimated by lenses set above the chamber and then divided into two paths with a beam splitter. One is detected with a photomultiplier, while the other is detected with an image intensifier. The trap was typically driven by rf voltages of 112 $V_{amp}$ at 22.2 MHz. For axial confinement, de voltages ranging from 3 V to 19 V were applied.

\section{Results}
In order to evaluate the magnetic-field gradient, we measured the Zeeman splitting of the $^2S_{1/2}$ - $^2D_{5/2}$ transition at several points along the $z$ direction in region B because the second-largest field gradient is expected without a large offset field of ${B_y}$ and ${B_z}$. After trapping a single Ca$^+$ ion in region B, a stray dc field was compensated for to minimize the excess micromotion of ions. In addition, we compensated for as much of the offset field of $B_x$ and $B_y$ as possible by using external coils attached outside the vacuum chamber. We then measured the splitting between two Zeeman components of the $^2S_{1/2}$ and $^2D_{5/2}$ transition at several points along the z axis.

\begin{figure}[htb!]
 \centering
 \includegraphics[width=80mm, trim=0mm 0mm 0mm 0mm, clip]{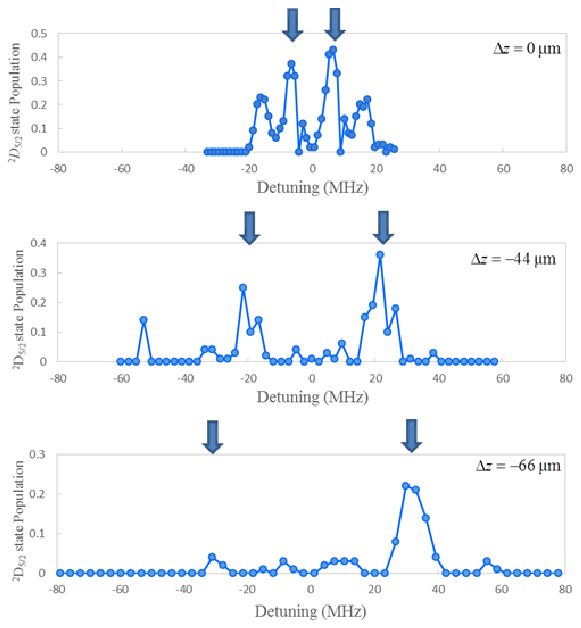}
 \caption{\label{figurefour}Spectra of the $^2S_{1/2}-^2D_{5/2}$ transition of a single $^{40}$Ca$^+$. $\Delta z$ represents the distance from the original ion position.}
\end{figure}

\begin{figure}[htb!]
 \centering
 \includegraphics[width=70mm, trim=00mm 0mm 0mm 0mm, clip]{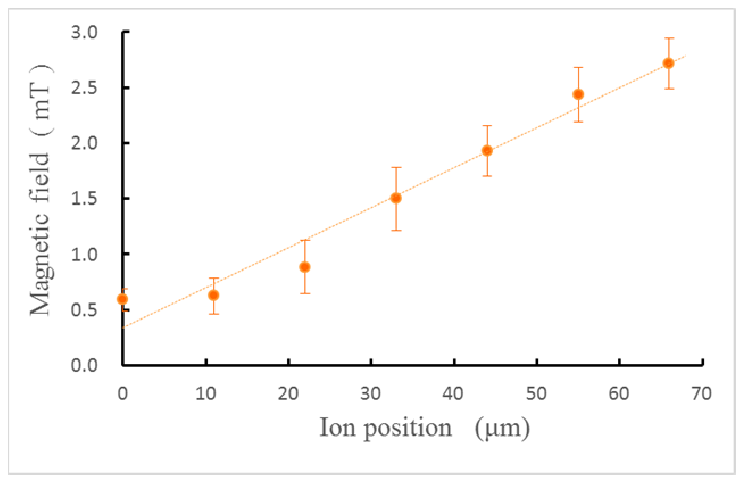}
 \caption{\label{figurefive} Magnitude of the magnetic field estimated from the Zeeman splitting at different positions of a single trapped ion.}
\end{figure}

Figure \ref{figurefour} shows the spectra of the $^2S_{1/2}-^2D_{5/2}$ transition at 729 nm obtained at different positions of the trap. Among the four $\Delta m_J=\pm 1$ transitions shown in the top panel in Fig.\ref{figurefour}, we focus on the spectra of two, $m_j = -1/2 \to m_j'= -3/2$ and $m_j = +1/2 \to m_j'= +3/2$, which are indicated by vertical arrows, to measure the Zeeman splitting of the $^2S_{1/2}$ state. The ion position is changed by the dc control voltages. Figure \ref{figurefive} shows the dependence of the magnetic-field magnitude estimated from the observed Zeeman splitting on the ion position. The error is mainly due to the width of the spectrum. By fitting a linear function, we estimated a magnetic-field gradient of 36 T/m.

　\smallskip 

　\smallskip

\section{Discussion}

We measured the field gradient in region B, where the second-largest magnetic-field gradient is expected. We also attempted to perform the experiments in region A, where the largest magnetic-field gradient is estimated; however, the fluorescence from Ca$^+$ ion could not be observed. We also attempted moving the ion from the region next to A to region A. We could shuttle the ion, However, it was not possible to monitor the ion image while the ion remained in region A. We consider that this is due to the large offset field originating from a shift between the trap electrode and the magnet layer, which causes a large detuning in the transition at 397 nm.  
 
we measured the size of the shift between the trap and the magnet layer by using a laser microscope and found that the size of the shift was on the order of several tens of micrometers. The adjustment of the ion position by dc voltages is possible; however, after the micromotion compensation \cite{micro}, we cannot apply an additional dc voltage freely. According to the calculation, a shift as high as 50 $\mu$m can produce an offset field of approximately 5 mT in the $z$ direction at $z=0$. It is difficult to cancel such a high offset field with our present external coils attached outside the vacuum chamber. To overcome the offset-field problem due to the shift between the trap electrode and the magnet layer, it is preferable to fabricate a magnet layer integrated with the trap electrode.

Other possible cause of excess offset magnetic field is that the magnitude of the magnetic field generated by each permanent magnet is uneven. To overcome this problem, we need to adjust the positional relation between the rf node and magnetic field. One possibility is to set the magnet layer on a movable stage inside the vacuum chamber that can be three-dimensionally moved with respect to the position of ions. Alternatively, it is possible to move the rf node by introducing an additional rf field to a trap electrode \cite{Osp2014,MIT2010,MIT2011}. The obtained field gradient of 36 T/m at a distance of approximately 350 $\mu$m from the magnet layer is in good agreement with the calculation. From this result, we can infer that a field gradient of 100 T/m is generated in region A.

\section{Conclusion}

We have demonstrated a surface-electrode trap with SmCo magnets arranged in the quadrupole configuration underneath the trap electrode. By utilizing the great advantage of the permanent magnets, which do not require any heat management and can be placed as close to ions as possible, a large magnetic-field gradient can be generated. A field gradient as large as 36 T/m has been estimated by measuring the Zeeman splitting of a single trapped $^{40}$Ca$^+$ ion at several positions, which is in good agreement with the calculation. From this result, we can infer that a field gradient of 100 T/m is generated in region A. A large field gradient is necessary to implement a sufficient state-dependent force, which is vital for quantum simulation and/or quantum gate operation using radio-frequency or microwave radiation. It is important to assemble the trap electrode and the magnet layer without any shift between them. The fabrication of a magnet layer integrated with the trap electrode would be a solution to overcome this problem. A field gradient on the order of 100 T/m would be possible with the presented method.

\section{Acknowledgments}

We are grateful to Yasuhiro Tokura for helpful advice on the numerical calculation of magnetic fields.


\section*{References}
\begin{thebibliography}{26}









  \bibitem{NIST1995} C. Monroe, D. M. Meekhof, B. E. King, W. M. Itano, and D. J. Wineland, Phys. Rev. Lett. {\bf 75}, 4714 (1995).



  \bibitem{Blatt2003cnot}F. Schmidt-Kaler, H. Haffner, M. Riebe, S. Gulde, G. P. T. Lancaster, T. Deuschle, C. Becher, C. F. Roos, J. Eschner, and R. Blatt, Nature, {\bf 422}, 408 (2003).


 \bibitem{Blatt2006cnot}M. Riebe, K. Kim, P. Schindler, T. Monz, P. O. Schmidt, T. K. K\"orber, W. H\"ansel, H. H\"affner, C. F. Roos, and R. Blatt, Phys. Rev. Lett. {\bf 97}, 220407 (2006).


 \bibitem{NIST1998twoent}Q. A. Turchette, C. S. Wood, B. E. King, C. J. Myatt, D. Leibfried, W. M. Itano, C. Monroe, and D. J. Wineland, Phys. Rev. Lett. {\bf 81}, 3631 (1998).

 


 \bibitem{NIST2000fourent} C. A. Sackett, D. Kielpinski, B. E. King, C. Langer, V. Meyer, C. J. Myatt, M. Rowe, Q. A. Turchette, W. M. Itano, D. J. Wineland, and C. Monroe, Nature, {\bf 404}, 256 (2000).
 



 \bibitem{Blatt2004}C. F. Roos, G. P. T. Lancaster, M. Riebe, H. H\"affner, W. H\"ansel, S. Gulde, C. Becher, J. Eschner, F. Schmidt-Kaler, and R. Blatt, Phys. Rev. Lett. {\bf 92}, 220402 (2004).


 \bibitem{Schaetz2008}A. Friedenauer, H. Schmitz, J. T. Glueckert, D. Porras, and T. Schtz, Nature Phys. {\bf 4}, 256 (2008).




 \bibitem{NISTspem2007}R. Ozeri, W. M. Itano, R. B. Blakestad, J. Britton, J. Chiaverini, J. D. Jost, C. Langer, D. Leibfried, R. Reichle, S. Seidelin, J. H. Wesenberg, and D. J. Wineland, Phys. Rev. A {\bf 75}, 042329 (2007).



 \bibitem{Wunde2001}F. Mintert and C. Wunderlich, Phys. Rev. Lett. {\bf 87}, 257904 (2001).  




 \bibitem{Chia2008}W. E. Chiaverini, J Lybarger  Jr., Phys. Rev. A {\bf 77}, 22324 (2008).



 \bibitem{Wunde2009}M. Johanning, A. Braun, N. Timoney, V. Elman, W. Neuhauser, and C. Wunderlich, Phys. Rev. Lett. {\bf 102}, 073004 (2009).


 \bibitem{Wunde2012}A. Khromova, C. Piltz, B. Scharfenberger, T. F. Gloger, M. Johanning, A. F. Var\'on, and C. Wunderlich, Phys. Rev. Lett. {\bf 108}, 220502 (2012).



 \bibitem{Wang2009} S. X. Wang, J. Labaziewicz, Y. Ge, R. Shewmon, and I. L. Chuang, Appl. Phys. Lett. {\bf 94}, 09410 (2009).



 \bibitem{Wunde2013}P. J. Kunert, D. Georgen, L. Bogunia, M. T. Baig, M. a. Baggash, M. Johanning, and C. Wunderlich, Appl. Phys. B {\bf 114} 27 (2013).  


 \bibitem{Osp2008}C. Ospelkaus, C. E. Langer, J. M. Amini, K. R. Brown, D. Leibfried, and D. J. Wineland, Phys. Rev. Lett. {\bf 101}, 090502 (2008).


 \bibitem{Osp2011}C. Ospelkaus, U. Warring, Y. Colombe, K. R. Brown, J. M. Amini, D. Leibfried, and D. J. Wineland, Nature {\bf 476}, 181 (2008).


 \bibitem{NISTtech2013}U. Warring, C. Ospelkaus, Y. Colombe, K. R. Brown, J. M. Amini, M. Carsjens, D. Leibfried, D. J. Wineland, Phys. Rev. A {\bf 87}, 013437 (2013).


 \bibitem{Osp2014}M. Carsjens, M. Kohnen, T. Dubielzig, C. spelkaus, Appl. Phys. B {\bf 114}, 243 (2014).



 \bibitem{Oxfordmw2014}D. P. L. Aude Craik, N. M. Linke, T. P. Harty, C. J. Ballance, D. M. Lucas, A. M. Steane, and D. T. C. Allcock, Appl. Phys. B {\bf 114}, 3 (2014).



 \bibitem{Radia}\texttt{http://www.esrf.eu/Accelerators/Groups\\/InsertionDevices/Software/Radia/}



 \bibitem{James1998}D. F. V. James, Appl. Phys. B {\bf 66}, 181 (1998).
 


 \bibitem{Roos2000}C. F. Roos, Ph.D. thesis, Universitat Innsbruck (2000).




 \bibitem{micro}Y. Ibaraki, U. Tanaka, and S. Urabe, Appl. Phys. B {\bf 105}, 219 (2011).


 \bibitem{MIT2010}T. H. Kim, P. F. Herskind, T. Kim, J. Kim, and I. L. Chuang, Phys. Rev. A {\bf 82}, 043412 (2010).

 \bibitem{MIT2011}T. H. Kim, P. F. Herskind, and I. L. Chuang, Appl. Phys. Lett. {\bf 98}, 214103 (2011).


\end{thebibliography}
\end{document}